\documentclass[amsmath,amssymb,11pt]{article}
\usepackage{amsfonts}
\usepackage{bm}
\date{}

\newcommand{\ot}{{\,\otimes\,}}
\newcommand{{\Cd}}{{\mathbb{C}^d}}

\newcommand{\sbalpha}{{\mbox{\scriptsize \boldmath $\alpha$}}}
\newcommand{\sbbeta}{{\mbox{\scriptsize \boldmath $\beta$}}}

\newcommand{\balpha}{{\mbox{ \boldmath $\alpha$}}}

\newcommand{\sbm}{{\mbox{\scriptsize \boldmath $m$}}}
\newcommand{\sbn}{{\mbox{\scriptsize \boldmath $n$}}}
\newcommand{\sbk}{{\mbox{\scriptsize \boldmath $k$}}}
\newcommand{\sbl}{{\mbox{\scriptsize \boldmath $l$}}}
\def\oper{{\mathchoice{\rm 1\mskip-4mu l}{\rm 1\mskip-4mu l}%
{\rm 1\mskip-4.5mu l}{\rm 1\mskip-5mu l}}}
\def\<{\langle}
\def\>{\rangle}
\newtheorem{theorem}{Theorem}
\newtheorem{corollary}{Corollary}
\newtheorem{proposition}{Proposition}
\newtheorem{example}{Example}

\begin{document}
\title{\textbf{A class of Bell diagonal states and entanglement witnesses}}

\author{Dariusz Chru\'sci\'nski, Andrzej Kossakowski and Krzysztof M{\l}odawski\\
Institute of Physics, Nicolaus Copernicus University \\
Grudzi\c{a}dzka 5/7, 87--100 Toru\'n, Poland \\ \\
Takashi Matsuoka \\
Tokyo University of Science, Suwa, Toyohira 5001\\  Chino City,
Nagano 391-0292, Japan }



\maketitle

\begin{abstract}
We analyze special class of bipartite states -- so called Bell
diagonal states. In particular we provide new examples of bound
entangled Bell diagonal states and construct the class of
entanglement witnesses diagonal in the magic basis.
\end{abstract}


\section{Introduction}

In recent years, due to the rapid development of quantum information
theory \cite{QIT} the necessity of classifying entangled states as a
physical resource is of primary importance. It is well known that it
is extremely hard to check whether a given density matrix describing
a quantum state of the composite system is separable or entangled.
There are several operational criteria which enable one to detect
quantum entanglement (see e.g. \cite{Horodecki-review} for the
recent review). The most famous Peres-Horodecki criterion  is based
on the partial transposition: if a state $\rho$ is separable then
its partial transposition $(\oper \ot \tau)\rho$ is positive. States
which are positive under partial transposition are called PPT
states. Clearly each separable state is necessarily PPT but the
converse is not true. We stress that it is easy to test wether a
given state is PPT, however, there is no general methods to
construct PPT states.


In \cite{CIRCULANT} (see also \cite{PPT-nasza}) we proposed a class
of bipartite states which is based on certain  decomposition of the
total Hilbert space $\mathbb{C}^d \ot \mathbb{C}^d$ into direct sum
of $d$-dimensional subspaces. This decomposition is controlled by
some cyclic property, that is, knowing one subspace, say $\Sigma_0$,
the remaining subspaces $\Sigma_1, \ldots, \Sigma_{d-1}$ are
uniquely determined by applying a cyclic shift to elements from
$\Sigma_0$. Now, we call a density matrix $\rho$ a {\it circulant
state} if $\rho$ is a convex combination of density matrices
supported on $\Sigma_\alpha$. The crucial observation is that a
partial transposition of the circulant state has again a circular
structure corresponding to another direct sum decomposition
$\widetilde{\Sigma}_0 \oplus \ldots \oplus
\widetilde{\Sigma}_{d-1}$.  Interestingly, also realignment \cite{R}
leaves the circulant structure invariant. This class was generalized
to multipartite systems \cite{CIRCULANT-multi}. Its separability
properties were analyzed in \cite{Art}.

The class of circulant states contains a subclass of states which
are diagonal in the basis of generalized Bell states in
$\mathbb{C}^d \ot \mathbb{C}^d$. The corresponding rank-1 projectors
define $d^2$--dimensional simplex known in the literature as {\em
magic simplex}. Several properties of Bell diagonal states were
analyzed \cite{Bell-1,Bell-1a,Bell-2,Bell-3,Bell-4}. In the present
paper we perform further studies of this special class of bipartite
states. In particular we provide new examples of bound entangled
Bell diagonal states and analyzed the class of entanglement
witnesses diagonal in the magic basis.

\section{Circulant states for two qudits}

Consider a class of states living in $\mathbb{C}^d \ot \mathbb{C}^d$
constructed as follows: let $\{e_0,\ldots,e_{d-1}\}$ denotes an
orthonormal basis in $\mathbb{C}^d$ and let $S : \mathbb{C}^d
\rightarrow \mathbb{C}^d$ be a shift operator defined as follows
\begin{equation}\label{}
    Se_k = e_{k+1} \ ,\ \ \ ({\rm mod}\ d) \ .
\end{equation}
One introduces $d$ $d$-dimensional subspaces in $\mathbb{C}^d \ot
\mathbb{C}^d$:
\begin{equation}\label{}
    \Sigma_0 = {\rm span}\{ e_0 \ot e_0, \ldots, e_{d-1} \ot e_{d-1}
    \} \ ,
\end{equation}
and
\begin{equation}\label{}
    \Sigma_n = (\mathbb{I} \ot S^n) \Sigma_0 \ , \ \ n=1,\ldots,d-1\
    .
\end{equation}
It is clear that $\Sigma_m$ and $\Sigma_n$ are mutually orthogonal
for $m \neq n$ and hence the collection $\{ \Sigma_0, \ldots ,
\Sigma_{d-1}\}$ defines direct sum decomposition of $\mathbb{C}^d
\ot \mathbb{C}^d$
\begin{equation}\label{}
\mathbb{C}^d \ot \mathbb{C}^d  = \Sigma_0 \oplus \ldots \oplus
\Sigma_{d-1}\ .
\end{equation}
To construct a circulant state corresponding to this decomposition
let us introduce $d$ positive $d \times d$ matrices $a^{(n)} = [
a^{(n)}_{ij}]\, ; \, n=0,1,\ldots,d-1$. Now, define $d$ positive
operators $\rho_n$ supported on ${\Sigma}_n$ via the following
formula
\begin{eqnarray}\label{}
\rho_n =  \sum_{i,j=0}^{d-1}\, a^{(n)}_{ij}\, e_{ij} \ot S^n\,
e_{ij}\, S^{*n} \ = \  \sum_{i,j=0}^{d-1}\, a^{(n)}_{ij}\, e_{ij}
\ot e_{i+n,j+n}\ .
\end{eqnarray}
Finally, we define the circulant density operator
\begin{equation}\label{}
    \rho = \rho_0 + \rho_1 + \ldots +
    \rho_{d-1}\ .
\end{equation}
 Normalization of
$\rho$, that is, ${\rm Tr}\, \rho =1$, is equivalent to the
following condition for matrices $a^{(n)}$
\[   \mbox{Tr}\, \left( a^{(0)} + a^{(1)} + \ldots +
a^{(d-1)} \right) = 1\ . \] The crucial property of circulant states
is based on the following observation \cite{CIRCULANT}: the
partially transposed circulant state $\rho$ displays similar
circulant structure, that is,
\begin{equation}\label{}
    ({\rm id} \ot \tau) \rho  = \widetilde{\rho}_0 \oplus \ldots
    \oplus \widetilde{\rho}_{d-1} \ ,
\end{equation}
where the operators $\widetilde{\rho}_n$ are supported on the new
collection of subspaces $\widetilde{\Sigma}_n$ which are defined as
follows:
\begin{equation}\label{}
    \widetilde{\Sigma}_0 = {\rm span}\{ e_0 \ot e_{\pi(0)}, e_1 \ot e_{\pi(1)}, \ldots, e_{d-1} \ot e_{\pi(d-1)}
    \} \ ,
\end{equation}
where $\pi$ is a permutation defined by
\begin{equation}\label{}
    \pi(k) = -k \ , \ \ \ ({\rm mod} \ d)\ .
\end{equation}
It means that
\[  \pi(0)=0, \ \pi(1)=d-1, \ \ldots , \pi(d-1) = 1 \ . \]
The remaining subspaces $\widetilde{\Sigma}_n$ are defined by a
cyclic shift
\begin{equation}\label{}
    \widetilde{\Sigma}_n = (\mathbb{I} \ot S^n) \widetilde{\Sigma}_0 \ , \ \ n=1,\ldots,d-1\
    .
\end{equation}
Again, the collection $\{ \widetilde{\Sigma}_0, \ldots ,
\widetilde{\Sigma}_{d-1}\}$ defines direct sum decomposition of
$\mathbb{C}^d \ot \mathbb{C}^d$
\begin{equation}\label{}
\mathbb{C}^d \ot \mathbb{C}^d  = \widetilde{\Sigma}_0 \oplus \ldots
\oplus \widetilde{\Sigma}_{d-1}\ .
\end{equation}
Moreover, operators $\widetilde{\rho}_n$ are defined as follows
\begin{eqnarray}\label{}
\widetilde{\rho}_n = \sum_{i,j=0}^{d-1}\, \widetilde{a}^{(n)}_{ij}\,
e_{ij} \ot S^n\, e_{\pi(i)\pi(j)}\, S^{*n} \ = \
\sum_{i,j=0}^{d-1}\, \widetilde{a}^{(n)}_{ij}\, e_{ij} \ot
e_{\pi(i)+n,\pi(j)+n}\ ,
\end{eqnarray}
with
\begin{equation}\label{a-tilde}
\widetilde{a}^{(n)} \, =\, \sum_{m=0}^{d-1}\, a^{(n+m)} \circ (\Pi
{S}^m)\ , \ \ \ \ \ \ \ \ (\mbox{mod $d$})\ ,
\end{equation}
where $\Pi$ is a permutation matrix corresponding to $\pi$, that is
\begin{equation}\label{}
    \Pi_{kl} = \delta_{k,\pi(l)}\ ,
\end{equation}
and $A \circ B$ denotes the Hadamard product of  $d\times d$
matrices $A$ and $B$.

\begin{theorem} A circulant state $\rho$ is PPT iff
$\ \widetilde{a}^{(n)} \geq 0\ $,
for $n=0,1,\ldots,d-1$.
\end{theorem}

It is clear that any circulant state $\rho$ gives rise to the
completely positive map $\Lambda : M_d(\mathbb{C}) \rightarrow
M_d(\mathbb{C})$ defined as follows
\begin{equation}\label{}
    \rho = ({\rm id} \ot \Lambda)P^+_d\ ,
\end{equation}
where $P^+_d$ denotes the maximally entangled state in $\mathbb{C}^d
\ot \mathbb{C}^d$, that is,
\begin{equation}\label{}
    P^+_d = \frac 1d \sum_{k,l=0}^{d-1} e_{kl} \ot e_{kl}\ .
\end{equation}
One easily finds the following formula for the action of $\Lambda$:
\begin{equation}\label{}
    \Lambda(e_{kl}) = \sum_{n=0}^{d-1} a^{(n)}_{kl} e_{k+n,l+n} \ .
\end{equation}
We call $\Lambda$ a circulant quantum channel  if $\Lambda$ is
unital, i.e. $\Lambda(\mathbb{I})=\mathbb{I}$. It implies the
following condition upon the collection of positive matrices
$a^{(n)}$:
\begin{equation}\label{}
    \sum_{k=0}^{d-1} \sum_{n=0}^{d-1} a^{(n)}_{kk} e_{k+n,k+n} = \mathbb{I} \ .
\end{equation}
Note, that a dual map $\Lambda^\#$ defined by ${\rm Tr}(\rho
\Lambda(X)) = {\rm Tr}(X \Lambda^\#(\rho))$, is defined as follows
\begin{equation}\label{}
    \Lambda^\#(e_{kl}) = \sum_{n=0}^{d-1} a^{(n)}_{lk} e_{k+n,l+n} \ ,
\end{equation}
i.e. it is defined by the collection of $a^{(n)T}$. It is well known
that if $\Lambda^\#$ is unital, then the original map $\Lambda$ is
trace preserving. Note, that in general $\Lambda$ is neither unital
nor trace preserving. In the next section we shall consider a
special class of circulant states which give rise to unital and
trace preserving circulant quantum channels.

\begin{example}
{\em A circulant state of 2 qubits has the following form
\begin{equation}\label{2C}
    \rho = \left( \begin{array}{cc|cc}
    a_{00} & \cdot & \cdot & a_{01} \\
    \cdot      & b_{00} & b_{01} & \cdot \\ \hline
    \cdot      & b_{10} & b_{11} & \cdot \\
    a_{10} & \cdot & \cdot & a_{11} \end{array} \right)\ .
\end{equation}
where for a more transparent presentation we introduced matrices $a
:= a^{(0)} \geq 0$ and $b := a^{(1)} \geq 0$.  Note, that a
circulant state (\ref{2C}) is usually called $X$-state in quantum
optics community \cite{X}. One easily finds for the partial
transposition
\begin{equation}\label{2C-T}
    \rho^\Gamma = \left( \begin{array}{cc|cc}
    \widetilde{a}_{00} & \cdot & \cdot & \widetilde{a}_{01} \\
    \cdot      & \widetilde{b}_{00} & \widetilde{b}_{01} & \cdot \\ \hline
    \cdot      & \widetilde{b}_{10} & \widetilde{b}_{11} & \cdot \\
    \widetilde{a}_{10} & \cdot & \cdot & \widetilde{a}_{11} \end{array} \right)\
    ,
\end{equation}
where the matrices $\widetilde{a} = [\widetilde{a}_{ij}]$ and
$\widetilde{b} = [\widetilde{b}_{ij}]$ read as follows
\begin{equation}\label{}
    \widetilde{a} = \left( \begin{array}{cc}
    a_{00} & b_{01} \\
    b_{10} & a_{11} \end{array} \right) \ , \ \ \ \
\widetilde{b} = \left( \begin{array}{cc}
    b_{00} & a_{01} \\
    a_{10} & b_{11} \end{array} \right)\ .
\end{equation}
Hence, $\rho$ defined in (\ref{2C}) is PPT iff
\begin{equation}\label{}
  \widetilde{a} \geq 0 \ \ \ \mbox{and} \ \ \ \widetilde{b} \geq 0
  \ ,
\end{equation}
and hence
\begin{eqnarray}\label{}
a_{00}a_{11} \geq |b_{01}|^2\ , \ \ \ \  b_{00}b_{11} \geq
|a_{01}|^2\ .
\end{eqnarray}
 }
\end{example}
\begin{example}
{\em A circulant state of 2 qutrits has the following form
\begin{equation}\label{3C}
 \hspace*{-.1cm}
  \rho = \left( \begin{array}{ccc|ccc|ccc}
    a_{00} & \cdot & \cdot & \cdot & a_{01} & \cdot & \cdot & \cdot & a_{02} \\
    \cdot& b_{00} & \cdot & \cdot & \cdot& b_{01} & b_{02} & \cdot & \cdot  \\
    \cdot& \cdot& c_{00} & c_{01} & \cdot & \cdot & \cdot & c_{02} &\cdot   \\ \hline
    \cdot & \cdot & c_{10} & c_{11} & \cdot & \cdot & \cdot & c_{12} & \cdot \\
    a_{10} & \cdot & \cdot & \cdot & a_{11} & \cdot & \cdot & \cdot & a_{12}  \\
    \cdot& b_{10} & \cdot & \cdot & \cdot & b_{11} & b_{12} & \cdot & \cdot  \\ \hline
    \cdot & b_{20} & \cdot & \cdot& \cdot & b_{21} & b_{22} & \cdot & \cdot \\
    \cdot& \cdot & c_{20} & c_{21} & \cdot& \cdot & \cdot & c_{22} & \cdot  \\
    a_{20} & \cdot& \cdot & \cdot & a_{21} & \cdot& \cdot & \cdot & a_{22}
     \end{array} \right)\ ,
\end{equation}
where $a := a^{(0)} \geq 0$, $b := a^{(1)} \geq 0$ and $c := a^{(2)}
\geq 0$. One easily finds for the partial transposition
\begin{equation}\label{3C-T}
 \hspace*{-.1cm}
  \rho^\Gamma = \left( \begin{array}{ccc|ccc|ccc}
    \widetilde{a}_{00} & \cdot & \cdot & \cdot &  \cdot & \widetilde{a}_{01} & \cdot &  \widetilde{a}_{02} & \cdot  \\
    \cdot& \widetilde{b}_{00} & \cdot & \widetilde{b}_{01} & \cdot & \cdot&   \cdot & \cdot & \widetilde{b}_{02}  \\
    \cdot& \cdot& \widetilde{c}_{00} &  \cdot & \widetilde{c}_{01} & \cdot & \widetilde{c}_{02} & \cdot & \cdot   \\ \hline
    \cdot & \widetilde{b}_{10} & \cdot &  \widetilde{b}_{11} & \cdot & \cdot & \cdot  & \cdot & \widetilde{b}_{12} \\
     \cdot & \cdot & \widetilde{c}_{10} & \cdot & \widetilde{c}_{11} & \cdot & \widetilde{c}_{12} & \cdot & \cdot   \\
    \widetilde{a}_{10} & \cdot & \cdot & \cdot & \cdot & \widetilde{a}_{11} & \cdot & \widetilde{a}_{12} & \cdot  \\ \hline
    \cdot &  \cdot & \widetilde{c}_{20} & \cdot & \widetilde{c}_{21} & \cdot &  \widetilde{c}_{22} & \cdot & \cdot \\
    \widetilde{a}_{20} & \cdot & \cdot & \cdot & \cdot & \widetilde{a}_{21} & \cdot & \widetilde{a}_{22} & \cdot  \\
     \cdot & \widetilde{b}_{20} & \cdot & \widetilde{b}_{21} & \cdot &  \cdot& \cdot & \cdot & \widetilde{b}_{22}
     \end{array} \right)\ ,
\end{equation}
where the matrices $\widetilde{a} = [\widetilde{a}_{ij}]$,
$\widetilde{b} = [\widetilde{b}_{ij}]$ and $\widetilde{c} =
[\widetilde{c}_{ij}]$ read as follows
\begin{eqnarray}\label{}
\widetilde{a} &=& \left( \begin{array}{ccc}
    a_{00} & c_{01} & b_{02} \\
    c_{10} & b_{11} & a_{12} \\
    b_{20} & a_{21} & c_{22} \end{array} \right) \ , \ \ \
\widetilde{b} \ = \left( \begin{array}{ccc}
    b_{00} & a_{01} & c_{02} \\
    a_{10} & c_{11} & b_{12} \\
    c_{20} & b_{21} & a_{22} \end{array} \right)\ ,
\ \ \
\widetilde{c} \ = \ \left( \begin{array}{ccc}
    c_{00} & b_{01} & a_{02} \\
    b_{10} & a_{11} & c_{12} \\
    a_{20} & c_{21} & b_{22} \end{array} \right)\ .
\end{eqnarray}

 }
\end{example}
For more examples see \cite{CIRCULANT}. Interestingly, circulant
structure is preserved under realignment.

\begin{proposition}
The realignment of the circulant bipartite operator
\begin{equation}\label{}
    A = \sum_{n,i,j=0}^{d-1} a^{(n)}_{ij} e_{ij} \ot e_{i+n,j+n} \ ,
\end{equation}
reads
\begin{equation}\label{}
  {\rm R}(A) = \sum_{n,i,j=0}^{d-1} R^{(n)}_{ij} e_{ij} \ot e_{i+n,j+n} \ ,
\end{equation}
where
\begin{equation}\label{}
    R^{(n)}_{ij} = a^{(j-i)}_{i+n,j}\ .
\end{equation}
\end{proposition}

\begin{example} {\em
The realignment of  $\rho$ defined in (\ref{2C}) leads to
\begin{equation}\label{2C}
    {\rm R}(\rho) = \left( \begin{array}{cc|cc}
    a_{00} & \cdot & \cdot & b_{00} \\
    \cdot      & a_{10} & b_{10} & \cdot \\ \hline
    \cdot      & b_{01} & a_{01} & \cdot \\
    b_{11} & \cdot & \cdot & a_{11} \end{array} \right)\
    .
\end{equation}
Hence in this case one has
\begin{equation}\label{}
    R^{(0)} = \left( \begin{array}{cc}
    a_{00} & b_{00} \\
    b_{11} & a_{11} \end{array} \right) \ , \ \ \ \
R^{(1)} = \left( \begin{array}{cc}
    a_{10} & b_{10} \\
    b_{01} & a_{01} \end{array} \right)\ .
\end{equation}

 }
\end{example}
\begin{example} {\em
The realignment of $\rho$ defined in (\ref{3C}) leads to the
circulant structure with
\begin{eqnarray}\label{}
R^{(0)}\ =\ \left( \begin{array}{ccc}
    a_{00} & b_{00} & c_{00} \\
    c_{11} & a_{11} & b_{11} \\
    b_{22} & c_{22} & a_{22} \end{array} \right) \ , \ \ \
R^{(1)} \ = \left( \begin{array}{ccc}
    a_{10} & b_{10} & c_{10} \\
    c_{21} & a_{21} & b_{21} \\
    b_{02} & c_{02} & a_{02} \end{array} \right)\ , \ \ \
R^{(2)} \ = \ \left( \begin{array}{ccc}
     a_{20} & b_{20} & c_{20} \\
    c_{01} & a_{01} & b_{01} \\
    b_{12} & c_{12} & a_{12}\end{array} \right)\ ,
\end{eqnarray}
and it my be easily generalized arbitrary dimension $d$.
 }
\end{example}

\section{Generalized Bell  diagonal states}

Consider now a simplex of Bell diagonal states
\cite{Bell-1,Bell-2,Bell-3,Bell-4} defined by
\begin{equation}\label{Bell}
    \rho = \sum_{m,n=0}^{d-1} p_{mn} P_{mn}\ ,
\end{equation}
where $p_{mn}\geq 0$, $\ \sum_{m,n}p_{mn}=1$ and
\begin{equation}\label{}
    P_{mn} = (\mathbb{I} \ot U_{mn}) \,P^+_d\, (\mathbb{I} \ot U_{mn}^\dagger)\ ,
\end{equation}
with $U_{mn}$ being the collection of $d^2$ unitary matrices defined
as follows
\begin{equation}\label{U_mn}
    U_{mn} e_k = \lambda^{mk} S^n e_k = \lambda^{mk} e_{k+n}\ ,
\end{equation}
with
\begin{equation}\label{}
    \lambda= e^{2\pi i/d} \ .
\end{equation}
The matrices $U_{mn}$ define an orthonormal basis in the space
$M_d(\mathbb{C})$ of complex $d \times d$ matrices. One easily shows
\begin{equation}\label{}
    {\rm Tr}(U_{mn} U_{rs}^\dagger) = d\, \delta_{mr} \delta_{ns} \ .
\end{equation}
Some authors \cite{Pittenger} call $U_{mn}$ generalized spin
matrices since for $d=2$ they reproduce standard Pauli matrices:
\begin{equation}\label{U-sigma}
    U_{00} = \mathbb{I}\ , \ U_{01} = \sigma_1\ , \ U_{10} = i
    \sigma _2\ , \ U_{11} = \sigma_3\ .
\end{equation}
Let us observe that Bell diagonal states (\ref{Bell}) are circulant
states in $\mathbb{C}^d \ot \mathbb{C}^d$. Indeed, maximally
entangled projectors $P_{mn}$ are supported on $\Sigma_n$, that is,
\begin{equation}\label{Pi_n}
    \Pi_n = P_{0n} + \ldots + P_{d-1,n} \ ,
\end{equation}
defines a projector onto $\Sigma_n$, i.e.
\begin{equation}\label{}
    \Sigma_n = \Pi_n ( \mathbb{C}^d \ot \mathbb{C}^d) \ .
\end{equation}
One easily shows that the corresponding matrices $a^{(n)}$ are given
by
\begin{equation}\label{}
    a^{(n)}= H D^{(n)} H^* \ ,
\end{equation}
where $H$ is a unitary $d\times d$ matrix defined by
\begin{equation}\label{}
    H_{kl} := \frac{1}{\sqrt{d}}\, \lambda^{kl} \ ,
\end{equation}
and $D^{(n)}$ is a collection of diagonal matrices defined by
\begin{equation}\label{}
    D^{(n)}_{kl} := p_{kn} \delta_{kl}\ .
\end{equation}
One has
\begin{equation}\label{}
    a^{(n)}_{kl} = \frac 1d \sum_{m=0}^{d-1} p_{mn}
    \lambda^{m(k-l)}\ ,
\end{equation}
and hence it defines a circulant matrix
\begin{equation}\label{}
   a^{(n)}_{kl} = f^{(n)}_{k-l}\ ,
\end{equation}
where the vector $f^{(n)}_m$ is the inverse of the discrete Fourier
transform of $p_{mn}$ ($n$ is fixed).

Consider now partial transposition of Bell diagonal states. One has
the following
\begin{theorem}
If $d$ is odd all matrices $\widetilde{a}^{(n)}$ are unitary
equivalent
\begin{equation}\label{}
\widetilde{a}^{(n)} = S^n \widetilde{a}^{(0)} S^{\dagger \, n}\ ,
\end{equation}
for $n=0,1,\ldots,d-1$. If $d$ is even one has two groups of unitary
equivalent matrices:
\begin{equation}\label{}
\widetilde{a}^{(2k)} = S^k \widetilde{a}^{(0)} S^{\dagger \, k}\ ,
\end{equation}
 and
\begin{equation}\label{}
\widetilde{a}^{(2k+1)} = S^{k} \widetilde{a}^{(1)} S^{\dagger \, k}\
,
\end{equation}
for $k=0,1,\ldots,d/2-1$.
\end{theorem}
Therefore

\begin{corollary}
Bell diagonal state is PPT if

\begin{itemize}

\item $\ \ \widetilde{a}^{(0)} \geq 0\,$ for $d$ odd,

\item  $\ \ \widetilde{a}^{(0)} \geq 0$ and $\  \,\widetilde{a}^{(1)} \geq 0$
for $d$ even.

\end{itemize}

\end{corollary}

The corresponding completely positive map $\Lambda : M_n(\mathbb{C})
\rightarrow M_n(\mathbb{C})$ is defined by the following Kraus
representation
\begin{equation}\label{Lambda-U}
    \Lambda(X) = \sum_{m,n=0}^{d-1} p_{mn} U_{mn} X U_{mn}^\dagger\
    ,
\end{equation}
where $p_{mn}\geq 0$ and $\sum_{m,n}p_{mn}=1$. One has
$\Lambda(\mathbb{I}) = \sum_{m,n} p_{mn}\, \mathbb{I} = \mathbb{I}$,
which proves that $\Lambda$ is unital. Note, that the dual map
\begin{equation}\label{Lambda-U*}
    \Lambda^\#(X) = \sum_{m,n=0}^{d-1} p_{mn} U_{mn}^\dagger X U_{mn}\
    ,
\end{equation}
is unital as well. Hence, $\Lambda$ defines unital and trace
preserving quantum channel (doubly stochastic completely positive
map).

\section{Special cases}

In this section we analyze special classes of Bell diagonal states.

\subsection{$d=2$}

For 2-qubit case one obtains the following density operator
\begin{eqnarray}\label{}
a^{(n)} =  \left( \begin{array}{cc}
    x_n & {y}_n \\
    y_n & x_n  \end{array} \right) \ ,
\end{eqnarray}
where
\begin{equation}\label{xn}
    x_n = \frac 12 ( p_{0n} + p_{1n} ) \ , \ \ \
    y_n = \frac 12 ( p_{0n} - p_{1n} ) \
    ,
\end{equation}
for $n=0,1$. The state is PPT if and only if
\begin{eqnarray}\label{}
    x_0^2  \geq  |y_1|^2\ , \ \ \ \
    x_1^2  \geq  |y_0|^2\ .
\end{eqnarray}
The above conditions imply well  known result that 2-qubit Bell
diagonal state is PPT (and hence separable) if and only if
\begin{equation}\label{1/2}
p_{mn} \leq \frac 12\ .
\end{equation}


\subsection{$d=3$}

For $d=3$ the Bell diagonal state is defined by the collection of 3
matrices
\begin{eqnarray}\label{}
a^{(n)} =  \left( \begin{array}{ccc}
    x_n & {z}_n & \overline{z}_n \\
    \overline{z}_n & x_n &  {z}_n \\
    {z}_n & \overline{z}_n & x_n \end{array} \right) \ , \ \ n=0,1,2\
    ,
\end{eqnarray}
where
\begin{equation}\label{xn}
    x_n = \frac 13 ( p_{0n} + p_{1n} + p_{2n}) \ ,
\end{equation}
and
\begin{equation}\label{zn}
    z_n = \frac 13 ( p_{0n} + \lambda p_{1n} + \overline{\lambda} p_{2n}) \
    .
\end{equation}
Now, the PPT condition reduces to the positivity of
$\widetilde{a}^{(0)}$
\begin{eqnarray}\label{}
\widetilde{a}^{(0)} =  \left( \begin{array}{ccc}
    x_0 & {z}_2 & \overline{z}_1 \\
    \overline{z}_2 & x_1 &  {z}_0 \\
    {z}_1 & \overline{z}_0 & x_2 \end{array} \right) \, \geq \, 0 \ ,
\end{eqnarray}
which is equivalent to the following conditions
\begin{equation}\label{C1}
    x_0x_1 \geq |z_2|^2  \ ,
\end{equation}
and
\begin{equation}\label{C2}
    x_0x_1x_2 + 2{\rm Re}\, z_0z_1z_2 \geq x_0 |z_0|^2 + x_1|z_1|^2 +
    x_2|z_2|^2\ .
\end{equation}
Hence, even for $d=3$ the PPT condition is by no means simple. It
might considerably simplify if we specify $x_n$ and $z_n$. Assume
for example that $z_0=0$. Then (\ref{C1})--(\ref{C2}) imply
\begin{equation}\label{}
    x_0x_1x_2 \geq x_1|z_1|^2 +
    x_2|z_2|^2\ .
\end{equation}

\subsection{$d=4$}

For $d=4$ the Bell diagonal state is defined by the collection of 4
matrices
\begin{eqnarray}\label{}
a^{(n)} =  \left( \begin{array}{cccc}
    x_n & {z}_n & y_n & \overline{z}_n \\
    \overline{z}_n & x_n &  {z}_n & y_n \\
    y_n &  \overline{z}_n & x_n & z_n \\
    z_n & y_n & \overline{z}_n & x_n \end{array} \right) \ , \ \
    n=0,1,2,3\ ,
\end{eqnarray}
where
\begin{eqnarray}\label{xn}
    x_n &=& \frac 14 ( p_{0n} + p_{1n} + p_{2n} + p_{3n}) \ ,
    \nonumber \\
    y_n &=& \frac 14 ( p_{0n} - p_{1n} + p_{2n} - p_{3n}) \ ,
     \\
     z_n &=& \frac 14 ( p_{0n} + ip_{1n} - p_{2n} -i p_{3n}) \ .
    \nonumber \
\end{eqnarray}
Bell diagonal state of two qutrits is PPT iff
\begin{eqnarray}\label{}
\widetilde{a}^{(0)} =  \left( \begin{array}{cccc}
    x_0 & {z}_4 & y_2 & \overline{z}_1 \\
    \overline{z}_4 & x_2 &  {z}_1 & y_0 \\
    y_2 &  \overline{z}_1 & x_0 & z_4 \\
    z_1 & y_0 & \overline{z}_4 & x_2 \end{array} \right)\, \geq \, 0 \ ,
\ \ \ {\rm and} \ \ \ \widetilde{a}^{(1)} =  \left(
\begin{array}{cccc}
    x_1 & {z}_0 & y_3 & \overline{z}_2 \\
    \overline{z}_0 & x_3 &  {z}_2 & y_1 \\
    y_3 &  \overline{z}_2 & x_1 & z_0 \\
    z_2 & y_1 & \overline{z}_0 & x_3 \end{array} \right) \, \geq \, 0 \
    .
\end{eqnarray}

\subsection{Special form of $p_{mn}$}

 Consider now special examples of Bell diagonal states by
specifying the structure of probability distribution $p_{mn}$. Let
\begin{equation}\label{}
    p_{mn} = \delta_{mk} \pi_n\ ,
\end{equation}
with $\pi_0 + \ldots + \pi_{d-1} = 1$. It gives rise to
\begin{equation}\label{n}
    \rho = \sum_{n=0}^{d-1} \pi_n P_{kn}\ .
\end{equation}
For example if $d=2$ and $k=0$ one obtains
\begin{equation}\label{}
    \rho = \frac 12 \left( \begin{array}{cc|cc}
    \pi_0 & \cdot & \cdot & \pi_0 \\
    \cdot      & \pi_1 & \pi_1 & \cdot \\ \hline
    \cdot      & \pi_1 & \pi_1 & \cdot \\
    \pi_0 & \cdot & \cdot & \pi_0 \end{array} \right)\ .
\end{equation}
This state is separable if and only if $\pi_0=\pi_1 = 1/2$. One
easily generalizes this observation as follows
\begin{proposition}
Bell diagonal state (\ref{n}) is separable if and only if
\begin{equation}\label{}
    \pi_0 = \ldots = \pi_{d-1} = \frac 1 d\ .
\end{equation}

\end{proposition}

Another characteristic class corresponds to
\begin{equation}\label{}
    p_{mn} = q_m p_n\ ,
\end{equation}
i.e. $p_{mn}$ represents the product distribution. One has
\begin{equation}\label{mn}
    \rho = \sum_{k,l=0}^{d-1} p_{kl} P_{kl} = \rho_0 \oplus \ldots
    \oplus \rho_{d-1} \ ,
\end{equation}
where
\begin{equation}\label{}
    \rho_n = p_n \sum_{m=0}^{d-1} q_m P_{mn} \ .
\end{equation}
Note, that matrices $a^{(n)}$ are related as follows
\begin{equation}\label{}
    a^{(n)} = p_n a\ ,
\end{equation}
where the matrix $a$ reads
\begin{equation}\label{}
    a_{kl} = \frac 1d \sum_{m=0}^{d-1} \lambda^{m(k-l)}\, q_m \ .
\end{equation}

\begin{proposition}
Bell diagonal state (\ref{mn}) is separable if and only if
\begin{equation}\label{}
    p_0 = \ldots = p_{d-1} = \frac 1 d\ .
\end{equation}

\end{proposition}

\subsection{Generalized lattice states}

Consider now a family of unitary operators acting on $N$ copies of
$\mathbb{C}^d$
\begin{equation}\label{}
    U_{\sbm\sbn} = U_{m_1n_1} \ot \ldots \ot U_{m_Nn_N}\ ,
\end{equation}
where $\mathbf{m} =(m_1,\ldots,m_N)$ and $\mathbf{n}
=(n_1,\ldots,n_N)$. It is clear that $U_{\sbm\sbn}$ defines a family
of $D^2=d^{2N}$ unitary operators in in $\mathbb{C}^{D} =
\mathbb{C}^{d \ot N}$. Note that
\begin{equation}\label{}
    {\rm Tr}(U_{\sbm\sbn} U^\dagger_{\sbk\sbl}) =
    D\, \delta_{\sbm\sbk}\delta_{\sbn\sbl}\ .
\end{equation}
Now, let $|\psi^+_D\>$ denote a maximally entangled state in
$\mathbb{C}^D \ot \mathbb{C}^D$ defined by
\begin{equation}\label{}
    \psi^+_D = \frac{1}{\sqrt{D}}\, \sum_{\sbk} e_{\sbk} \ot
    e_{\sbk}  \ ,
\end{equation}
where
\begin{equation}\label{}
    e_{\sbk} = e_{k_1} \ot \ldots \ot e_{k_N} \ .
\end{equation}
One defines a family of maximally entangled states by
\begin{equation}\label{}
 |\psi_{\sbm\sbn}\> = (\mathbb{I} \ot U_{\sbm\sbn})|\psi^+_D\>  \ .
\end{equation}
These states are parameterized a  point $(\mathbf{m},\mathbf{n})$ in
the $N$-dimensional lattice $L^{(d)}_{(N)}$ consisting of $D^2$
points. Now, a generalized lattice state is defined by a collection
of points from $L^{(d)}_{(N)}$: for any subset $I \subset
L^{(d)}_{(N)}$ one defines
\begin{equation}\label{}
    \rho_I = \frac{1}{|I|}\, \sum_{(\sbm,\sbn) \in I} P_{\sbm\sbn}\ ,
\end{equation}
where $P_{\sbm\sbn} = |\psi_{\sbm\sbn}\>\<\psi_{\sbm\sbn}|$ and
$|I|$ stands for the cardinality of $I$. Clearly, $1 \leq |I| \leq
|L^{(d)}_{(N)}| = D^2$. Let us observe that the above construction
generalized a class of lattice states presented in
\cite{Benatti-OSID,Piani,Benatti-JMP}. Lattice states of Benatti et.
al. correspond to $d=2$. In this case $U_{mn}$ are defined in terms
of Pauli matrices (see formula (\ref{U-sigma})).

\section{Bound entangled Bell diagonal states}

\subsection{Two qutrits}

Consider the following family of Bell diagonal  states
\begin{equation}\label{eps}
    \rho_\varepsilon = N_\varepsilon (P_{00} + \varepsilon \Pi_1 +
    \varepsilon^{-1} \Pi_2 ) \ ,\ \ \ \varepsilon > 0 \ ,
\end{equation}
where the projectors $\Pi_k$ are defined in (\ref{Pi_n}) and the
normalization factor reads
\begin{equation}\label{}
    N_\varepsilon = \frac{1}{1+ \varepsilon + \varepsilon^{-1}}\ .
\end{equation}
It corresponds to (cf. formulae (\ref{xn}) and (\ref{zn}))
\begin{equation}\label{}
    x_0 = \frac{N_\varepsilon}{3}\ , \ \  x_1 = \frac{N_\varepsilon}{3}\, \varepsilon\ , \
    \ x_2 = \frac{N_\varepsilon}{3}\, \varepsilon^{-1}\ ,
\end{equation}
and
\begin{equation}\label{}
    z_0 = \frac{N_\varepsilon}{3}\ , \ \ z_1=z_2=0\ .
\end{equation}
Hence, conditions (\ref{C1})--(\ref{C2}) are trivially satisfied
showing that (\ref{eps}) defines a family of PPT states. Now, it is
well known \cite{Jur09} that $\rho_\varepsilon$ is separable if and
only if $\varepsilon=1$. Hence, for $\varepsilon \neq 1$ it defines
a family of bound entangled state in $\mathbb{C}^3 \ot
\mathbb{C}^3$. The entanglement of $\rho_\varepsilon$ can be
detected by using a realignment criterion \cite{R}. In the next
section we show that it can be detected also by the Bell diagonal
entanglement witness. Note, that asymptotically
\begin{equation}\label{}
    \lim_{\varepsilon \rightarrow 0} \rho_\varepsilon = \frac 13\, \Pi_2\
    ,\ \
    \lim_{\varepsilon \rightarrow \infty} \rho_\varepsilon = \frac 13\, \Pi_1\
    ,
\end{equation}
that is, one obtains separable states defined by normalized
separable projectors onto $\Sigma_2$ and $\Sigma_1$, respectively.

\subsection{Two qudits}

Consider a family of states in $\mathbb{C}^d\otimes\mathbb{C}^d$
defined by  \cite{Ha,How}
\begin{equation}\label{}
\rho_\gamma= \frac 1 N_\gamma \sum_{i,j=0}^{d-1} e_{ij} \otimes
A_{ij}^\gamma\,,
\end{equation}
where $d\times d$ matrices
\begin{equation}\label{}
A_{ij}^\gamma = \left\{\begin{array}{cl}
e_{ij} &\mbox{for}\;i\neq j\,,\\[1ex]
 e_{00} +a_\gamma e_{11} +\sum_{\ell=2}^{d-2} e_{\ell\ell}+b_\gamma e_{d-1,d-1} &\mbox{for}\;i=j=0\,,\\[1ex]
S^{j-1}A^\gamma_{00}S^{\dagger j-1} &\mbox{for}\;i=j\neq 1
\end{array}\right.
\end{equation}
with
\begin{equation}\label{}
a_\gamma=\frac{1}{d}(\gamma^2+d-1)\,,\qquad
b_\gamma=\frac{1}{d}(\gamma^{-2}+d-1)\ ,
\end{equation}
and the normalization factor reads
\begin{equation}\label{}
N_\gamma = d^2-2+\gamma^2+\gamma^{-2} \ .
\end{equation}
It gives the following spectral decomposition
\begin{equation}\label{}
    \rho_\gamma = \frac{1}{N_\gamma} \left( dP_{00} + a_\gamma \Pi_1 +
    \sum_{\ell=2}^{d-2} \Pi_\ell + b_\gamma \Pi_{d-1} \right) \ .
\end{equation}
 In particular for
$d=3$ one obtains the following matrix representation:
\begin{equation}
  \rho_\gamma\ = \  \frac{1}{N_\gamma}\left( \begin{array}{ccc|ccc|ccc}
 1 &  \cdot& \cdot& \cdot& 1 & \cdot& \cdot& \cdot & 1 \\
 \cdot& a_\gamma &\cdot& \cdot& \cdot& \cdot& \cdot& \cdot& \cdot\\
 \cdot& \cdot& b_\gamma & \cdot& \cdot& \cdot& \cdot& \cdot& \cdot  \\ \hline
 \cdot& \cdot& \cdot& b_\gamma & \cdot& \cdot& \cdot& \cdot& \cdot \\
 1 & \cdot& \cdot& \cdot& 1 & \cdot& \cdot& \cdot& 1 \\
 \cdot& \cdot& \cdot& \cdot& \cdot& a_\gamma& \cdot & \cdot& \cdot  \\ \hline
 \cdot& \cdot& \cdot& \cdot& \cdot& \cdot & a_\gamma& \cdot& \cdot \\
 \cdot & \cdot& \cdot& \cdot& \cdot& \cdot& \cdot& b_\gamma& \cdot \\
 1& \cdot& \cdot& \cdot& 1 & \cdot& \cdot& \cdot& 1
  \end{array} \right)\ = \ \frac{1}{N_\gamma} (3P_{00} + a_\gamma
  \Pi_1 + b_\gamma \Pi_2 ) \ ,
\end{equation}
with $a_\gamma = \frac 13 (\gamma^2 + 2)\,$, $b_\gamma = \frac 13
(\gamma^{-2} + 2)\,$  and the normalization factor $N_\gamma = 7 +
\gamma^2 + \gamma^{-2}$.

\section{Bell diagonal entanglement witnesses}  \label{S-EW}

Interestingly many well known entanglement witnesses displaying
circulant structure are Bell diagonal. It is well known that any
entanglement witness $W$ can be represented as a difference $W = W_+
- W_-$, where both $W_+$ and $W_-$ are semi-positive operators in
$\mathcal{B}(\mathbb{C}^d \ot \mathbb{C}^d)$. However, there is no
general method to recognize that $W$ defined by $W_+ - W_-$ is
indeed an EW. An interesting class of such witnesses may be
constructed using their spectral properties \cite{CMP,Gniewko}. Let
$\psi_\alpha$ ($\alpha =0,1,\ldots,d^2-1$) be an orthonormal basis
in $\mathbb{C}^d \ot \mathbb{C}^d$ and denote by $P_\alpha$ the
corresponding projector $P_\alpha = |\psi_\alpha\>\<\psi_\alpha|$.
Now, take $d^2$  semi-positive numbers $\lambda_\alpha \geq 0$ such
that $\lambda_\alpha$ is strictly positive for $\alpha > L$, and
define
\begin{equation}\label{}
    W_- = \sum_{\alpha=0}^L \lambda_\alpha P_\alpha\ , \ \ \ \
W_+ = \sum_{\alpha=L}^{d^2-1} \lambda_\alpha P_\alpha\ ,
\end{equation}
where $L$ is an arbitrary integer $0<L<d^2-1$. This construction
guarantees that $W_+$ is strictly positive and all zero modes and
strictly negative eigenvalues of $W$ are encoded into $W_-$.
Consider normalized vector $\psi \in \mathbb{C}^d \ot \mathbb{C}^d$
and let $\ s_1(\psi) \geq \ldots \geq s_{d}(\psi)\ $ denote its
Schmidt coefficients. For any $1 \leq k \leq d$ one defines $k$-norm
of $\psi$ by the following formula
\begin{equation}\label{}
    || \psi ||^2_k = \sum_{j=1}^k s^2_j(\psi)\ .
\end{equation}
It is clear that
\begin{equation}\label{}
    ||\psi ||_1 \leq ||\psi||_2 \leq \ldots \leq ||\psi ||_d \ .
\end{equation}
Note that $||\psi||_1$ gives the maximal Schmidt coefficient of
$\psi$, whereas due to the normalization, $||\psi||^2_d =
\<\psi|\psi\> =1$.  One proves  \cite{CMP}  the following

\begin{theorem} Let $\ \sum_{\alpha=0}^{L-1} ||\psi_\alpha||^2_{k} < 1\,$.
If the following spectral conditions are satisfied
\begin{equation} \label{T1}
     \lambda_\alpha \geq \mu_k\ , \ \ \
    \alpha=L,\ldots,d^2-1\ ,
\end{equation}
where
\begin{equation}  \label{mu-l}
    \mu_\ell := \frac{\sum_{\alpha=0}^{L-1} \lambda_\alpha
    ||\psi_\alpha||^2_\ell}{1-\sum_{\alpha=0}^{L-1}
    ||\psi_\alpha||^2_\ell}\ ,
\end{equation}
then $W$ is an $k$-EW. If moreover $\ \sum_{\alpha=1}^L
||\psi_\alpha||^2_{k+1} < 1\ $ and
\begin{equation}\label{T2}
    \mu_{k+1} > \lambda_\alpha\ , \ \ \
    \alpha=L,\ldots,d^2-1\ ,
\end{equation}
then $W$ being $k$-EW is not $(k+1)$-EW.
\end{theorem}
Let us observe that if $\psi$ is maximally entangled then
\begin{equation}\label{}
    ||\psi||^2_k = \frac{k}{d} \ .
\end{equation}
Consider, therefore, the family of Bell diagonal states $\psi_{mn}$.
On has the following

\begin{corollary}
If $L < d$ and
\begin{equation}\label{}
\lambda_\alpha \geq \mu_1\ , \ \ \
    \alpha=L,\ldots,d^2-1\ ,
\end{equation}
with $\mu_1 = \frac{1}{d- L} \sum_{\alpha=0}^{L-1} \lambda_\alpha\,
$, then $W = W_+ - W_-$ defines Bell diagonal entanglement witness.
\end{corollary}

\begin{example}
{\em Consider well known entanglement witness in $d=2$ represented
by the flip operator
\begin{equation}\label{Flip}
    F =  \left( \begin{array}{cc|cc}
    1 & \cdot & \cdot & \cdot \\
    \cdot      & \cdot & 1 & \cdot \\ \hline
    \cdot      & 1 & \cdot & \cdot \\
    \cdot & \cdot & \cdot & 1 \end{array} \right)\ .
\end{equation}
Note that
\begin{equation}\label{}
    F = P_{00} + P_{10} + P_{01} - P_{11} \ ,
\end{equation}
which proves that $F$ is Bell diagonal and possesses single negative
eigenvalue.
 }
\end{example}
\begin{example}
{\em A family of EWs in $\mathbb{C}^3 \ot \mathbb{C}^3$ defined by
\cite{Cho-Kye}
\begin{equation}\label{}
  W[a,b,c]\, =\,  \left( \begin{array}{ccc|ccc|ccc}
    a & \cdot & \cdot & \cdot & -1 & \cdot & \cdot & \cdot & -1 \\
    \cdot& b & \cdot & \cdot & \cdot& \cdot & \cdot & \cdot & \cdot  \\
    \cdot& \cdot & c & \cdot & \cdot & \cdot & \cdot & \cdot &\cdot   \\ \hline
    \cdot & \cdot & \cdot & c & \cdot & \cdot & \cdot & \cdot & \cdot \\
    -1 & \cdot & \cdot & \cdot & a & \cdot & \cdot & \cdot & -1  \\
    \cdot& \cdot & \cdot & \cdot & \cdot & b & \cdot & \cdot & \cdot  \\ \hline
    \cdot & \cdot & \cdot & \cdot& \cdot & \cdot & b & \cdot & \cdot \\
    \cdot& \cdot & \cdot & \cdot & \cdot& \cdot & \cdot & c & \cdot  \\
    -1 & \cdot& \cdot & \cdot & -1 & \cdot& \cdot & \cdot & a
     \end{array} \right)\ ,
\end{equation}
with $a,b,c\geq 0$. Necessary and sufficient conditions for
$W[a,b,c]$ to be an EW are
\begin{enumerate}
\item $0 \leq a < 2\ $,
\item $ a+b+c \geq 2\ $,
\item if $a \leq 1\ $, then $ \ bc \geq (1-a)^2$.
\end{enumerate}
 A family $W[a,b,c]$ generalizes celebrated Choi
indecomposable witness corresponding to $a=b=1$ and $c=0$. One finds
the following spectral representation
\begin{eqnarray}\label{}
W[a,b,c] = (a-2)P_{00} + (a+1)(P_{10} + P_{20}) + b\Pi_{1} + c
\Pi_{2} \ ,
\end{eqnarray}
which shows that $W[a,b,c]$ is Bell diagonal with a single negative
eigenvalue `$a-2$'. }
\end{example}
\begin{example}
{\em Consider a family of EWs defined by \cite{How}
\begin{equation}\label{W-lm}
 W_{\lambda,\mu}\ = \  \left( \begin{array}{ccc|ccc|ccc}
 1 &  \cdot& \cdot& \cdot& -1 & \cdot& \cdot& \cdot & -1 \\
 \cdot& 1+\mu &\cdot& \cdot& \cdot& \mu& \mu& \cdot& \cdot\\
 \cdot& \cdot & \lambda & \lambda & \cdot& \cdot& \cdot& \lambda & \cdot  \\ \hline
 \cdot& \cdot & \lambda & \lambda & \cdot& \cdot& \cdot& \lambda & \cdot \\
 -1 & \cdot& \cdot& \cdot& 1 & \cdot& \cdot& \cdot& -1 \\
 \cdot& \mu & \cdot& \cdot& \cdot& 1+\mu & \mu & \cdot& \cdot  \\ \hline
 \cdot& \mu & \cdot& \cdot& \cdot& \mu & 1+\mu & \cdot& \cdot \\
 \cdot & \cdot& \lambda & \lambda & \cdot& \cdot& \cdot& \lambda & \cdot \\
 -1& \cdot& \cdot& \cdot& -1 & \cdot& \cdot& \cdot& 1
  \end{array} \right)\ ,
\end{equation}
with $\lambda,\mu \geq  0$. Note that $W_{0,0} = W[1,1,0]$. One
obtains for the spectral decomposition
\begin{equation}\label{}
W_{\lambda,\mu} = - 3P_{00} + 2 \Pi_0 + \Pi_1 + 3\mu P_{01} +
3\lambda P_{02}\ .
\end{equation}
Let $\gamma > 0$. One shows \cite{How} that for
\begin{equation}\label{lambda}
 \lambda < \frac{1-\gamma^2}{2 + \gamma^{-2}}\ ,
\end{equation}
and
\begin{equation}\label{}
 \mu < \frac{1-\gamma^2 - \lambda(2 + \gamma^{-2})}{2+\gamma^2}\ ,
\end{equation}
$W_{\lambda,\mu}$ defines an indecomposable EW due to ${\rm
Tr}(W_{\lambda,\mu}\rho_\gamma)<0$.
 }
\end{example}
\begin{example}
{\em Entanglement witness corresponding to the reduction map
$\Lambda(X) = \mathbb{I} {\rm Tr}X - X$ in $M_d(\mathbb{C})$. One
has
\begin{eqnarray}\label{}
    W = \frac 1d\, \mathbb{I} \ot \mathbb{I} - P^+_d  =
 \frac 1d \sum_{k,l=0}^{d-1} P_{kl} - P_{00}  \ ,
\end{eqnarray}
which shows that $W$ is Bell diagonal with a single negative
eigenvalue $(1-d)/d$. }
\end{example}
\begin{example} {\em A family of EWs in $\mathbb{C}^d \ot \mathbb{C}^d$ defined by
 \cite{Ha,How}
\begin{equation}\label{}
W_{d,k}=\sum_{i,j=0}^{d-1} e_{ij} \otimes X_{ij}^{d,k}\ ,
\end{equation}
where the $d\times d$ matrices $X_{ij}^{d,k}$ are defined as
\begin{equation}\label{}
X_{ij}^{d,k}=\left\{\begin{array}{cl}
 (d-k-1)e_{ii} +\sum_{\ell=1}^k S^\ell\, e_{ii}\,S^\ell &\mbox{for}\;i=j\,,\\[1ex]
- e_{ij} &\mbox{for}\;i\neq j\,,
\end{array}\right.
\end{equation}
It is well known \cite{Ha} that  $W_{d,k}$ defines an indecomposable
EW for $k=1,2,\ldots,d-2$. For $k=d-1$ it reproduces the witness
corresponding to the reduction map. Note that for $d=3$ and $k=1$ it
reproduces $W[a,b,c]$ with $a=b=1$ and $c=0$. On easily finds the
following spectral representation
\begin{equation}\label{}
     W_{d,k} = (d-k)\Pi_0 + \sum_{\ell=1}^k \Pi_k - dP_{00}\ ,
\end{equation}
showing that $W_{d,k}$ is Bell diagonal and the single negative
eigenvalue corresponds to the maximally entangled state $P_{00}$. }
\end{example}

\section{Conclusions}

We analyzed a class of bipartite circulant states which are diagonal
with respect to generalized Bell (magic) basis. Such states are
characterized by an elegant symmetry which considerably simplifies
their analysis. We analyzed several examples of bound entangled
states and provided corresponding entanglement witnesses which are
Bell diagonal.

\section*{Acknowledgments}

This work was partially supported by the Polish Ministry of Science
and Higher Education Grant No 3004/B/H03/2007/33.

\end{document}